\documentclass[aps,preprint,superscriptaddress,showpacs]{revtex4}
\usepackage{amsmath,bm}
\usepackage{graphicx}

\begin{document}

\title{Fano Resonance and Orbital Filtering in Multiply Connected Carbon Nanotubes} 
\author{Gunn Kim}
\author{Sang Bong Lee}
\author{Tae-Suk Kim}
\author{Jisoon Ihm}
\email[corresponding author. E-mail:\ ]{jihm@snu.ac.kr}
\affiliation{School of Physics, Seoul National University, Seoul 151-747, Korea}
\date{\today }

\begin{abstract}

We investigate the electron transport in multiply connected metallic carbon nanotubes
within the Landauer-B\"{u}ttiker formalism.
Quasibound states coupled to the incident $\pi^{*}$ states 
give rise to energy levels of different widths depending on the coupling strength.
In particular, donorlike states originating from heptagonal rings 
are found to give a very narrow level.
Interference between broad and narrow levels
produces Fano-type resonant backscattering as well as resonant tunneling.
Over a significantly wide energy range, almost perfect suppression of
the conduction of $\pi^{*}$ electrons occurs,
which may be regarded as filtering of particular electrons ($\pi$-pass filter).

\end{abstract}
\maketitle

\section{Introduction}
In the past decade, carbon nanotubes (CNTs) have been studied extensively
because of their unconventional properties from both the fundamental research and
application point of view~\cite{Dresselhaus}.
For the application to nanoscale electronic devices, researchers have fabricated
various forms of CNTs to engineer their physical properties.
$sp^2$-bonded carbon nanostructures have been found to coalesce
by electron beam irradiation at high temperature
and new morphologies such as X- and T-shaped junctions have been produced
~\cite{Terrones1,Terrones2,Bandow}. 
These developments offer interesting opportunities to study phase-coherent transport
in novel geometries. Thus far, phase-coherent transport has been most actively studied 
in the semiconductor heterostructures by employing the Aharonov-Bohm (AB) interferometer
\cite{closering,openring1,openring2,openring3,topology} with an embedded quantum dot in one arm.
The measured phase in the open geometry (as in double-slit experiments) 
is featured with a smooth increase at the resonant tunneling and an abrupt jump 
by 180$^\circ$ at transmission zeroes.
In the closed or unitary geometry, the conductance  
generally exhibits the Fano line shape~\cite{kobayashi}. 
The Fano effect usually arises from the coherent interference
between a narrow localized level (quantum dot) and 
a continuum energy spectrum (the arm without a quantum dot)~\cite{Fano}.
The Fano effect has also been identified in the direct transport through
a single quantum dot~\cite{fanodotexp1,fanodotexp2}.
The CNT is excellent  
for observing phase coherence phenomena and 
there are some experimental signatures of the Fano effect in the 
CNTs~\cite{JKim,WYi,ZZhang}, 
though the detailed mechanism is not fully understood.

In this paper, we report transport properties of
a multiply connected carbon nanotube (MCCN) structure as shown in Fig. 1,
where a single tube is branched off into two smaller arms and then they merge into one~\cite{Grimm}.
We have studied conductance variation systematically by changing the length of the arms.
Both $\pi$-bonding and $\pi^*$($\pi$ anti)-bonding electron transport channels 
show resonant tunneling through discrete energy levels(DELs) in the finite arms. 
The width of the resonant tunneling peaks in the $\pi$ channel is
broad and the transmission probability is fairly uniform as a function of energy.
The $\pi^*$ channel, on the other hand, has more interesting structure of either broad  
or narrow resonant tunneling as will be shown later.  
Coherent interference between a very broad level (extending over to nearby levels) and
its narrow neighboring level is evident in the asymmetric 
Fano-type line shapes of the $\pi^*$ transport channel 
and the corresponding transmission probability is featured with both zero and unity.
In particular, we find a nearly perfect suppression of the $\pi^*$ transmission 
in a certain finite range of energy. This mechanism can be regarded 
as electron filtering of a particular wave function character ($\pi$-pass filter), 
and such a nonequilibrium distribution may be maintained over the relaxation length 
of the electron phase beyond the arm region.

\section{Computational Details}
Our model system, the MCCN, comprises two leads of semi-infinite metallic (10,10) CNTs 
and two arms of finite (5,5) tubes (effectively a resistive region) of the same length in between.
This structure may be thought as two Y-junctions faced to each other,
possibly produced experimentally by electron irradiation on nanotubes.
Six heptagons are contained in each of two junction regions 
where the (10,10) tube and two (5,5) tubes are joined. 
The length of the arm region is represented by the number 
($L$) of periodic units of the armchair CNT as shown in Fig. 1. 
This structure possesses the mirror symmetry 
with respect to three planes. Two symmetry planes $\sigma_p$ and $\sigma_c$,  
which are, respectively, perpendicular to and containing the (10,10) tube axis, 
are indicated in Fig. 1. 
The third one is the plane of the figure containing the tube axis 
(not indicated in Fig. 1).
The electronic structure is described by the single $\pi$-electron tight-binding 
Hamiltonian
\begin{equation}
H = V_{pp\pi} \sum_{<i,j>}\left (a^{\dagger}_{i}a_{j} + h.c.\right),
\end{equation}
where $<i,j>$ denote the nearest neighbor pairs, 
the hopping integral $V_{pp\pi}=-2.66$ eV~\cite{Blase} and
the on-site energy is set to zero.

It is well known that the $\pi$ and $\pi^*$ bands are crossing at the Fermi level ($E_F$) in the armchair CNTs
as shown in Fig. 2. A $\pi$-bonding state consists of $p_z$ orbitals  having the same phase all across
the circumference of the tube.
In a $\pi$-antibonding ($\pi^*$) state, on the other hand, the constituent $p_z$ orbitals have alternating
signs (180$^{\circ}$ out of phase) along the circumference of the tube.
$\pi$ and $\pi^*$ states are usually admixed when defects are introduced into the perfectly periodic nanotubes. 
However, $\pi$ and $\pi^*$ transport channels remain unmixed in our MCCN
because of the $\sigma_{c}$ symmetry of the structure.
%
%
Only if $\pi$ and $\pi^*$ states are unmixed, the effect of interference among 
different levels becomes pronounced and dramatic (e.g., complete suppression
of transmission of a level) as described below.
Furthermore, because two arms are of the same length, we can focus on
other effects than the traditional AB interference due to the path length difference.
In our scattering-state approach, we solve the Schr\"odinger equation
for the whole system by matching the solutions of the tight-binding 
Hamiltonian in the arm region with that of the lead region at each interface.
For given incoming electron waves from the left lead as initial conditions, 
we obtain the transmitted electrons emerging on the right lead. 
Conductance is obtained from the Landauer-B\"{u}ttiker formula,
$G(E)=G_{0} {\rm Tr}({\bf t^{\dagger}t})$, where $G_0$ is the conductance quantum 
($=2e^2/h$) and ${\bf t}$ is the transmission matrix~\cite{Lan-Butt}.
We do not include here extra contact resistance that may arise between external
metallic leads and the MCCN.

\section{Results and Discussion}
The conductance as a function of the incident electron energy is displayed 
in Fig. 3 for the cases of $L=$~5, 9 and 12. 
To examine the structure of the conductance in detail, 
total conductance is decomposed into two nonmixing contributions of 
$\pi$ and $\pi^*$ channels, $G = G_0 (T_{\pi} + T_{\pi^*})$, where
$T_{\pi}$ and $T_{\pi^*}$ are the transmission probabilities of the two.
We find the following features common to various $L$'s.
(i) $T_{\pi}$ has peaks of magnitude one always and varies very slowly as a function of energy. 
(ii) $T_{\pi^*}$ has both broad and narrow peaks of magnitude one always,
and the line shape is highly asymmetric, especially for narrow peaks. 
(iii) $T_{\pi}$ has no zeroes within the interested energy window while 
$T_{\pi^*}$ is featured with zeroes near the narrow asymmetric peaks.

Since the wave function of a $\pi$ state does not have phase variation 
in the circumferential direction 
(i.e., the same sign over the cross-section of the tube) irrespective of $n$ in ($n$,$n$) tubes,
the $\pi$ wave function in the lead part ((10,10) tube) has  
almost perfect match with that in the arms (two (5,5) tubes),
and the transmission is close to unity in a wide range of energy.
In contrast, the conduction of $\pi^*$ electrons 
is more complicated because the wave function in the lead part cannot
match that of the two-arms region due to the phase variation
along the circumferential direction as mentioned before. 
The conducting behavior may be explained by a close examination of the resonant tunneling 
through the DELs 
which are accommodated in the arms of (5,5) tubes.
Each junction area acts as an effective scattering center 
for $\pi^*$ electrons.
Thus the MCCN structure may be regarded as a double-wall quantum well and 
we expect the energy quantization by confinement to a finite region of the arms.
With increasing length $L$ of (5,5) tubes, the spacing between peaks becomes smaller.
Peaks for the $\pi^*$ states in Fig. 3 show this trend unambiguously. 
($\pi$ electrons also experience resonant tunneling 
via DELs as exemplified in the series of peaks of magnitude unity.
However, their line width is so broad that the variation in the conductance is small 
and uninteresting.) The spectrum of DELs can be roughly estimated from 
the linear dispersion~\cite{Mintmire} near $E_F$ for the ideal armchair nanotube,
\begin{eqnarray}
|E(k)| &=& \frac{a \sqrt{3}}{2} |V_{pp\pi}| |k-k_{F}|, 
\end{eqnarray}
where $a$ (= 2.46 \AA) is the lattice constant and
$k_{F}$ ($=2\pi/3a$) is the Fermi wave vector.
Approximate values of the spectrum of DELs are found from the quantization condition 
$kd=m\pi$, where $d$ is the effective length of the arm and $m$ is positive integer.
For instance, if we took $d=La~(L=13)$ for $k=m\pi/d$, 
the DELs near $E_F$ would be $-$0.93, $-$0.37, 0.19, and 0.74 eV.
A better fitting to the exact numerical calculation in Fig. 4(a) 
could be obtained if we assumed a longer effective length
$d \approx (L+1.5)a$, in which case we would have $-$0.82, $-$0.32, 0.18, and 0.68 eV.
This result is reasonable since the electrons localized in the central ring region are
somewhat spread out beyond the arms as seen, for instance, in the last picture 
of Fig. 4(b). This simple model based on the linear dispersion, of course, has 
a limitation of obtaining, incorrectly, the same energy separations only.

Coupling to the left and right leads 
results in the finite lifetime (line broadening) of the DELs.   
To understand the variation of the line width from peak to peak,
we further investigate the structure of the wave functions at the resonant peaks.
Besides the linear $\pi$ and $\pi^*$ bands crossing at $E_F$, 
there exist a conduction band at 0.8 eV and above, and a valence band at $-$0.8 eV 
and below in the (10,10) tube.
Because of the presence of the heptagonal carbon rings at the junction,
donorlike states tend to be produced according to well-known 
H\"uckel's ($4n+2$) rule for stability~\cite{Solomons}.
The cycloheptatrienyl cation is a molecular example of the heptagonal ring 
following H\"uckel's rule for $n=1$.
The localized impurity (donor) state is most pronounced 
if the energy level is close to (and below) aforementioned conduction band minimum 
at 0.8 eV.
Such localized donorlike states have in fact been found 
in other defects (nitrogen impurity and Stone-Wales defect) of the nanotube 
as well~\cite{HChoiprl,HKim}.
The peak C in Fig. 4 (or sharp peaks closest to 0.8 eV in Fig. 3)
corresponds to this state and has a narrow line shape
owing to the relatively well-localized (long lifetime) character of such an origin.
The DELs further away from the conduction band minimum are less localized and thus broader.

We also compare the wave functions around the peaks and the nodes of the conductance curve.
The wave functions at peaks of $T_{\pi^*}$ have large amplitude in two arms
so that incident electrons are in resonance with the states in the region.
In contrast, the wave functions of the $\pi^{*}$ band 
at the U-shaped valley (e.g., at 0.46 eV in Fig. 4(b))
are depleted inside two arms, which means an effective
disconnection of two leads and the suppression of the electron conduction.
Unlike in the $\pi$ electron case, the phase variation across the cross-section of the tube 
is very rapid in the $\pi^*$ state and it is very difficult to match the wave function 
at the junction boundary for a general value of energy, hence substantial reduction 
in conductance.

So far, we have accounted for the interesting phenomenon of 
line width variations from level to level. 
Such a variation leads to even more profound consequences
on the structure of the transmission probability as we shall describe below.
In the 1D mesoscopic systems such as the double-barrier quantum wells,
if all the DELs are coupled to the continuum of lead states
with almost the same coupling strength and 
the relative signs of couplings to two leads are alternating from level to level, 
the transmission zero does not occur \cite{Silva,tskim}.
However, when one energy level is coupled by far more strongly to the leads
and acts like a broad band interfering with other narrow levels, 
the transmission zero may take place \cite{tskim}.
Our system falls in the latter case.
As the energy of the DELs is increased, 
the number of nodes along the two arms increases one by one 
and in turn the relative signs of the wave functions 
at two junctions are oscillating for DELs.
The DELs in the arm region
are coupled to the left and right (10,10) tube leads with the same coupling strength but with variable signs.
The signs of the coupling constants at both junctions are determined by the parity of the wave functions with respect to
the reflection symmetry plane, $\sigma_p$. For an even-parity state, 
the signs of coupling constants at two junctions
are the same, but they are opposite for an odd-parity state. 
This alternating parity under $\sigma_p$ symmetry operation is clearly visible 
in Fig. 4(b) for DELs at $-$0.32 (even), 0.18 (odd), and 0.59 eV (even).
In the presence of the $\sigma_p$ symmetry 
(plus the time-reversal symmetry which obviously exists in the absence of 
applied magnetic fields),
it is well known that the perfect transmission ($T_{\pi^*}=1$) is allowed to occur
at all resonant peaks in the system~\cite{HWLee}.
On the other hand, the transmission zero is generally not allowed 
in the energy range between one level
and its nearest neighbor level interfering each other~\cite{tskim}.
It is only when a broad level extends beyond the nearest neighbor narrow level
that a transmission zero occurs at the energy between its nearest 
and {\it next nearest neighbor} narrow level.

To confirm this argument, we fit the curve of $T_{\pi^*}$ for $L=13$ with the multilevel 
generalization of the Green's function formula~\cite{Caroli,Hershfield,Meir}.
The expression for $T$ contains, as parameters, DELs ($E_i$) and
coupling constants to right($V_{Ri}$) and left leads($V_{Li}$)~\cite{tskim,Caroli,Hershfield,Meir}. 
\begin{eqnarray}
T(\varepsilon) &=& 4{\rm Tr}\{\Gamma_L D^r(\varepsilon) \Gamma_R D^a(\varepsilon)\},
\end{eqnarray}
\begin{eqnarray}
D^{r,a}(\varepsilon) &=& [\varepsilon I - E \pm i \Gamma]^{-1},
\end{eqnarray}
where $\Gamma_{L,R}$ is the coupling matrix between the resistive and lead region
on the left or right, $D^{r,a}$ is the retarded or advanced Green's function, $E_{ij}=E_i \delta_{ij}$, 
$\Gamma_{ij}=\Gamma_{Lij} + \Gamma_{Rij}=\pi N_L V_{iL} V_{Lj} + \pi N_R V_{iR} V_{Rj}$
and $N_{L,R}$ is the density of states in the lead on the left or right ($N_{L}=N_{R}$ in the present case).
We use seven DELs and their diagonal components of $\Gamma$.
Values of DELs ($\Gamma_{ii}$) in eV are as follows: 
$-$1.330(0.139), $-$0.859(0.194), $-$0.300(0.132), 0.178(0.025),
0.584(0.013), 1.100(0.183), and 1.700(0.250).
The conductance curve in much wider energy range than the one presented in Fig. 4 
has been fitted for more complete results.
As mentioned above, relative signs of the wave function 
at two junctions alternate from one level to the next so that we set $V_{Ri}=\lambda V_{Li}$,
where $\lambda$ is 1 ($-$1) for the even (odd) parity state with respect to $\sigma_p$.
$\Gamma$ represents the broadening of a particular level and results in the finite energy width 
of the conductance peak.
In Fig. 4, A, B, and C are the broadened peaks associated with respective DELs.
At a given energy, contributions from tails of two or more broadened peaks overlap and give rise to
an asymmetric line shape.
For instance, between A and B (i.e., the right-hand side of A and left-hand side of B), 
conduction paths interfere constructively and no transmission zero occurs. 
However, on the right-hand side of B, A and B interfere destructively 
(since the energy is greater than both A and B) and produce a transmission zero.
Since peak C is very narrow, 
its constructive interference on B (in the energy range between B and C) 
is much less than the destructive interference from A and unable to change the result. 
In fact, there are two transmission zeroes between B and C,
one nearer to B which is obtained as above and the other nearer to C 
which is obtained by the destructive interference between C 
and the next broad peak above C (not shown).
Here we emphasize that the transmission zeroes result from the interference between the abstract 
conduction paths through the energy levels in the Hilbert space. 
The curve fitting with Eq.(3) is practically indistinguishable 
from the $\pi^*$-channel conductance of our MCCN, 
indicating that the simple model with the multilevel transmission formula  
is valid for explaining the Fano resonance structures and the transmission zeroes.
We also observe that, in a certain finite energy range,
the conductance of $\pi^*$ electrons is almost zero
while that of $\pi$ electrons is not reduced appreciably.
For instance, the transmitted electrons of $\pi^*$ character for $L=5$ in Fig. 3(a)
constitute less than 8\% of the total transmitted electrons in the energy range
between 0 and 0.5 eV.
This may be regarded as orbital filtering of a particular wave function character ($\pi$-pass filter).
This result can be used for a further fundamental research on the orbital character 
in nanostructures or an application to current switching in nanotubes.

\section{Summary}
We have carried out tight-binding calculations of conductance 
of multiply connected metallic CNTs with the mirror symmetry.
The $\pi$ channel is well-conducting and has very broad resonance peaks. 
The $\pi^*$ channel exhibits Fano resonance structures. 
The Fano effect originates from the interference
between broad and narrow electronic levels in the arm region.
It is shown that the present system transmits predominantly $\pi$ 
over $\pi^*$ electrons in a certain energy range.

This work was supported by the CNNC of Sungkyunkwan University, the MOST through the NSTP
(grant No. M1-0213-04-0001), the Grant No. 1999-2-114-005-5 from the KOSEF,
the Korea Research Foundation Grant (KRF-2003-042-C00038), the Samsung SDI-SNU Display
Innovation Program, and the Supercomputing Application Support Program of the KISTI.

\newpage
\begin{center}
\LARGE{[Figure Captions]}
\end{center}

Figure 1: Model of MCCN with L=10. $\sigma_{p}$ and $\sigma_{c}$ denote the planes of mirror symmetry.

Figure 2: Band structure of the (10,10) armchair carbon nanotube near the Fermi level, $E_{F}(=0)$.

Figure 3: (color). Conductance as a function of the incident energy {\it E}~~for the MCCNs. 
The Fermi level is set to zero. (a) L=5, (b) L=9, and (c) L=12.

Figure 4: (color). Conductance and wavefunctions for L=13. 
In (a), the full numerical calculation for $\pi^*$ (solid red line) is indistinguishable 
from the fit with the multiple Fano resonance formula (dashed red line). 
In (b), the red and blue colors indicate $\pm$ signs and the size of the spheres is the amplitude of the wavefunction.

\newpage

\begin{figure}[t]
  \centering
  \includegraphics[width=14cm]{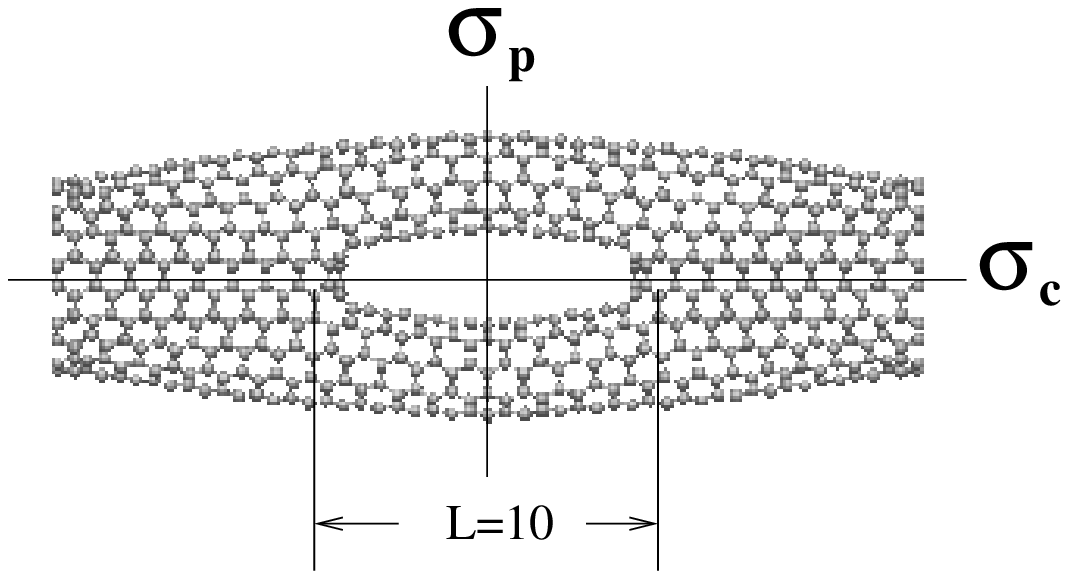}
  \label{model}
\end{figure}
\begin{center}
\LARGE{Figure 1}

\LARGE{G. Kim, S.B. Lee, T.-S. Kim and J. Ihm}
\end{center}

\newpage
\begin{figure}[t]
  \centering
  \includegraphics[width=14cm]{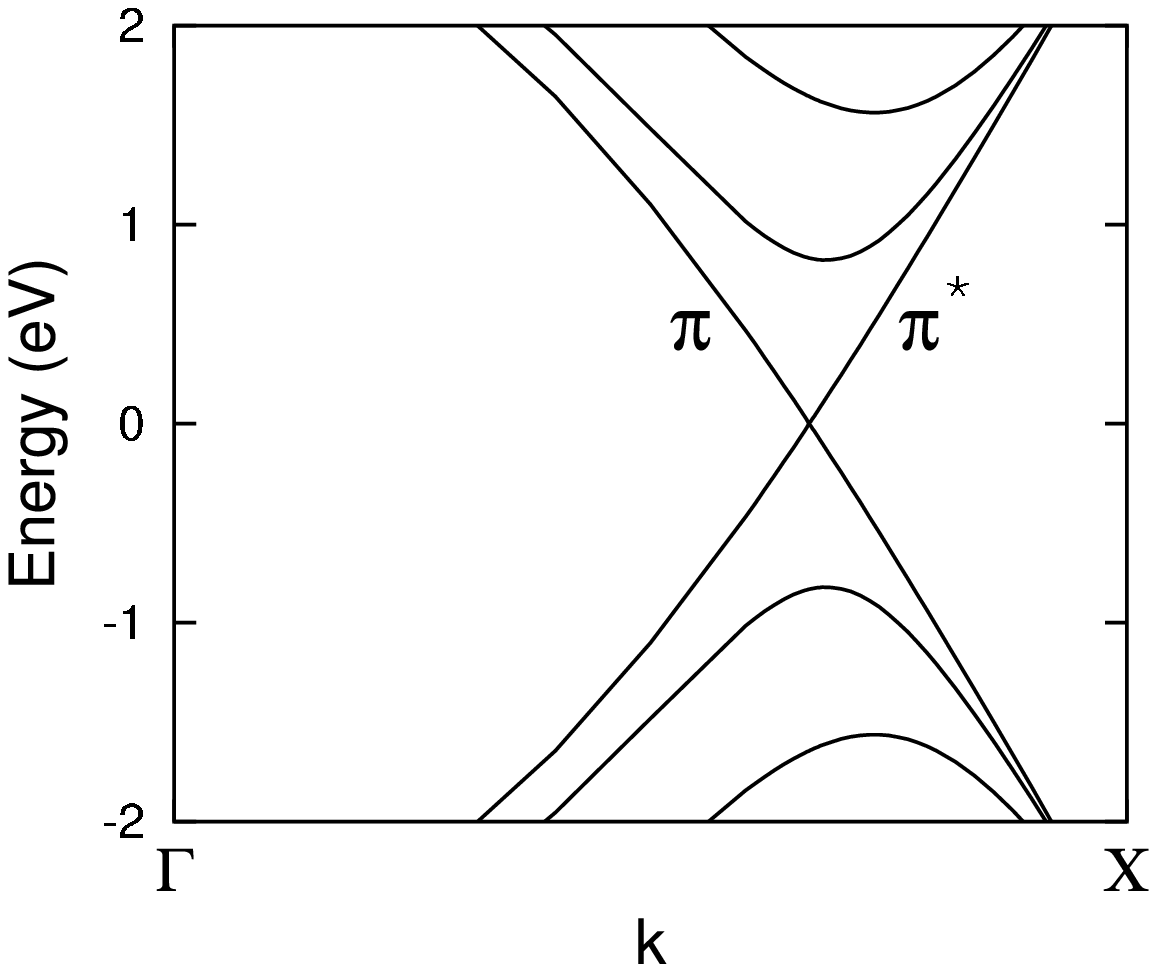}
  \label{bandstructure}
\end{figure}
\begin{center}
\LARGE{Figure 2}

\LARGE{G. Kim, S.B. Lee, T.-S. Kim and J. Ihm}
\end{center}

\newpage
\begin{figure}[t]
  \centering
  \includegraphics[width=14cm]{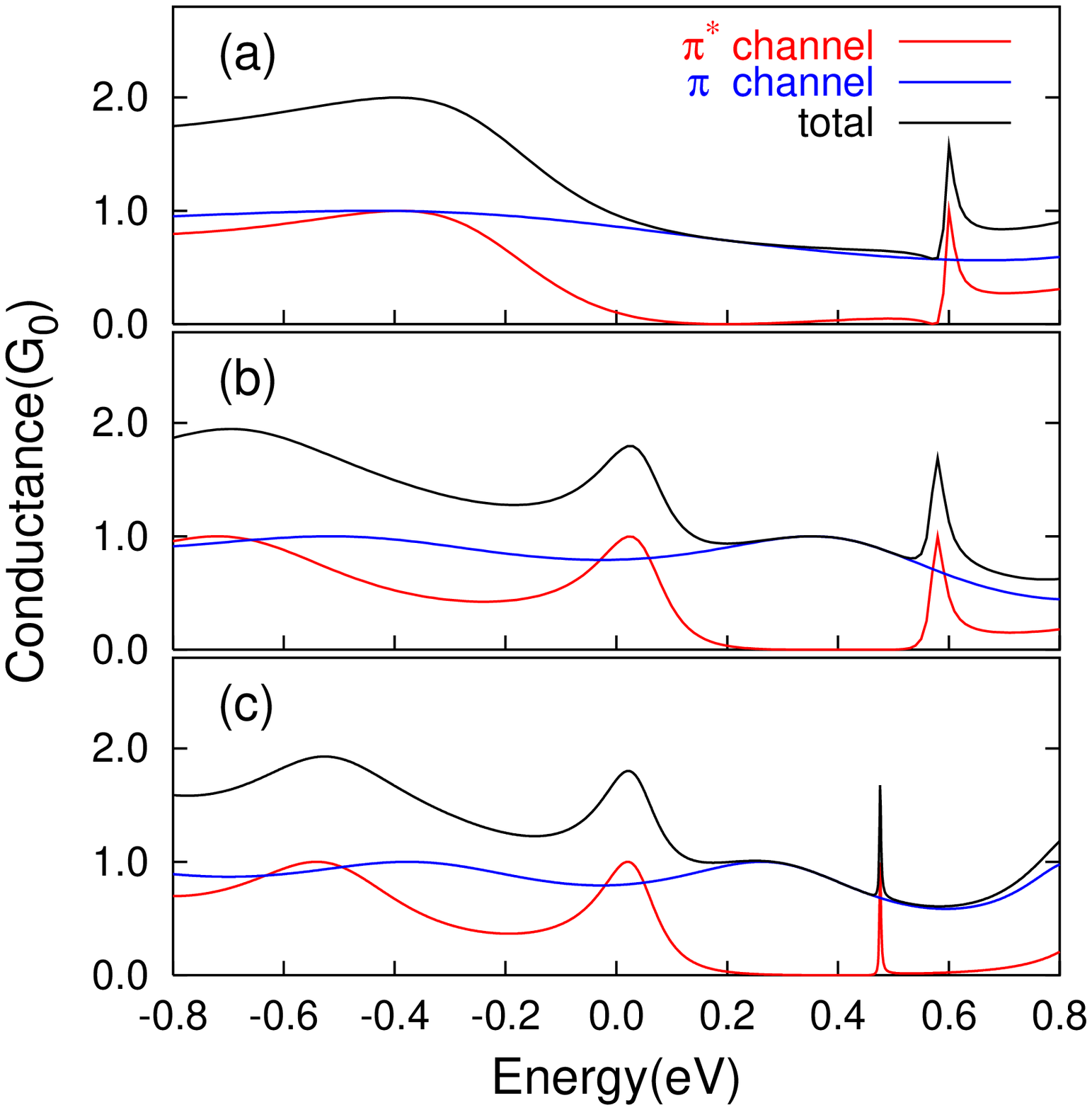}
  \label{conductance}
\end{figure}
\begin{center}
\LARGE{Figure 3}

\LARGE{G. Kim, S.B. Lee, T.-S. Kim and J. Ihm}
\end{center}

\newpage
\begin{figure}[t]
  \centering
  \includegraphics[width=14cm]{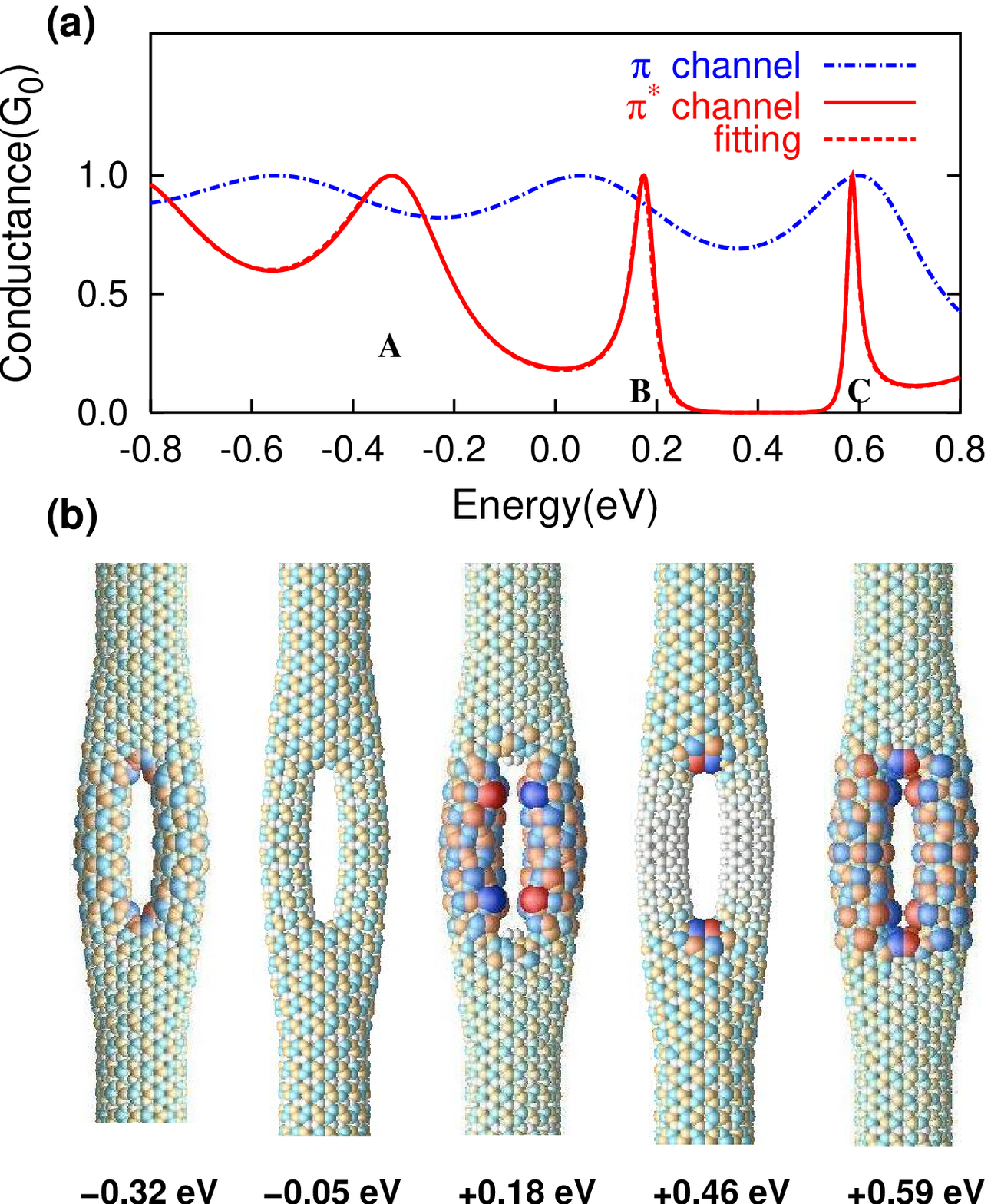}
  \label{wavefunction}
\end{figure}
\begin{center}
\LARGE{Figure 4}

\LARGE{G. Kim, S.B. Lee, T.-S. Kim and J. Ihm}
\end{center}

\end{document}